**Waveform Selective Surfaces**

*Hiroki Wakatsuchi,\* Jiang Long, and Daniel F. Sievenpiper*

((Optional Dedication))

Prof. H. Wakatsuchi
Center for Innovative Young Researchers
Nagoya Institute of Technology
Gokiso-cho, Showa, Nagoya, Aichi, 466-8555, Japan
Department of Electrical and Mechanical Engineering
Nagoya Institute of Technology, Gokiso-cho, Showa, Nagoya, Aichi, 466-8555, Japan
E-mail: wakatsuchi.hiroki@nitech.ac.jp
Dr. J. Long
Skyworks Solution Inc., San Jose, California, 95134, USA
Prof. D. F. Sievenpiper
Applied Electromagnetics Group, Electrical and Computer Engineering Department,
University of California, San Diego, La Jolla, California 92093, USA

((Abstract text))

The role of frequency is very important in electromagnetics since it may significantly change how a material interacts with an incident wave if the frequency spectrum varies. Here, we demonstrate a new kind of microwave window that has the unique property of controlling transmission and reflection based not only on the frequency of an incoming wave but also on the waveform or pulse width. This surface can preferentially pass or reject different kinds of signals, such as short pulses or continuous waves, even if they occur at the same frequency.

Such a structure can be used, for example, to allow long communication signals to pass through, while rejecting short radar pulses in the same frequency band. It is related to the classic frequency selective surface, but adds the new dimension of waveform selectivity, which is only possible by introducing nonlinear electronics into the surface. Thus, our study is expected to provide new solutions to both fundamental and applied electromagnetic issues ranging from traditional antenna design and wireless communications to emerging areas such as cloaking, perfect lenses, and wavefront shaping.

((Main Text Paragraphs))

The response of a material to an incident wave changes as the frequency component of the wave changes. Conversely, materials are generally incapable of distinguishing between different waves at the same frequency unless these materials exhibit, for example, nonlinearity[1], anisotropy[2–4], or other dependences on thermal heating[5,6] or light intensity[7]. In this study, we show that pulse width can be used as an additional degree of freedom - specifically, we use metasurfaces[8] composed of several circuit elements to control the scattering of an incident wave at the same frequency depending on the waveform (**Figure 1**a). In addition, we demonstrate that these waveform-selective metasurfaces are readily tuned externally or internally to realize more complex scattering responses. Moreover, our structures have the potential to exploit a conventional design drawback or physical limitation of an application as design flexibility. Note that our study shows a simple design methodology to control not only absorption[9,10] but also a wide range of scattering parameters through the use of periodic conductor effects. In addition, circuit-level effects enhance the design flexibility in waveform-selective responses. Our structures are related to the classic frequency selective surface (FSS)[11], but add the new dimension of waveform selectivity.

Such unusual electromagnetic responses are obtained using conventional periodic structures with several circuit elements, including sets of four diodes that play the role of diode bridges[9]. Thus, the energy of an incoming wave is converted into an infinite set of frequency spectra, although most of the energy is at zero frequency (see Table S1 in Supporting Information). Within each diode bridge, we paired a resistor with either a capacitor or an inductor (Figure 1b, c). This allowed the control of rectified electric charges through the time-domain response of these paired circuit elements, as a capacitor gradually increases its electric potential and thereby reduces the number of incoming electric charges, while an inductor lowers its electromotive force, allowing more electric charges to enter[10]. In essence, these metasurfaces respond to how long an incident wave continues but not to the difference in the frequency spectrum, which appears due to the discontinuity of the waveform at the beginning and end (see Supporting Information Figure S1 for more details).

These circuit characteristics were first integrated with two types of periodic structures, specifically, cut-wire and slit structures, which exhibit resonant mechanisms at a particular frequency (Figure 1d, g and Supporting Information Figure S3a, b). When designed complementarily using Babinet's principle[12], these structures enhance either reflection (in the case of cut-wire structures) or transmission (in the case of slit structures) at the same frequency. In our study, however, they are not designed to be complementary to each other to improve waveform-selective responses but are instead used as specific examples that control the level of these fundamental scattering parameters. These structures contain either of the above-mentioned circuit elements, i.e., Figure 1b or c, between conductor edges where the electric field intensifies due to each resonant phenomenon (see the circled numbers in Figure 1d, g and the insets of Figure 1e, h). For this reason, the resonant mechanism is associated with the rectification process in response to the incoming wave pulse width.

These structures were numerically tested by using an electromagnetic/circuit co-simulation solver (see the Methods section). We found that, for a cut-wire structure composed of a capacitor-based circuit (i.e., Figure 1b), the reflecting performance is limited for a 50-ns short pulse at 4.0 GHz (Figure 1e) because induced electric charges are effectively rectified and the intrinsic resonance of the cut-wire structure is disrupted. However, this rectification process is mitigated for a continuous wave (CW) signal so that the resonance emerges again at 4.0 GHz.

Similarly, a strong level of transmission is observed in Figure 1h for a CW at 4.2 GHz, while this performance is suppressed in the case of a short pulse at the same frequency.

These waveform-selective resonant mechanisms were also designed using an inductor-based circuit (i.e., Figure 1c) for a different type of waveform. In this case, a short pulse is not effectively rectified due to the electromotive force of the inductor. Thus, a cut-wire structure strongly reflects the incoming wave as a conventional structure in Figure 1f. However, a sufficiently long pulse or a CW reduces the electromotive force and is thus effectively rectified by the diodes. This eventually leads to a suppression of the intrinsic resonant mechanism of the cut-wire structure.

A third method for controlling fundamental electromagnetic scattering is shown in Figure 1i and j, where an incident wave is reflected back with a different polarization by deploying a symmetric L-shaped conductor as a periodic unit together with a ground plane (GND) (see the inset of Figure 1j and Supporting Information Figure S3c for its small signal response)[13]. This structure contains capacitor-based circuits and thus exhibits a reduced polarization change for a short pulse at 3.5 GHz compared to a CW at the same frequency.

Importantly, the waveform-selective response can be tuned by replacing some of the circuit elements. For instance, the periodic unit cell depicted in **Figure 2**a has a conducting square patch and a ground plane above and underneath a substrate, respectively, and thus exhibits waveform-selective absorption, if a capacitor-based circuit is introduced (Supporting Information Figure S6a)[9]. In the diode bridge circuit, the role of the resistor is to determine the magnitude of the reflected power at the steady state (Supporting Information Figure S6b), since it varies the amount of electric charges to be released for energy dissipation at the steady state. To realize further waveform-selective responses through the use of a single circuit element, the resistor can be replaced with a junction gate field-effect transistor (jFET) (Figure 2b). This is because the gate voltage $V_g$ varies the amount of the current between the drain and source (Supporting Information Figure S7), and therefore the effective resistance between these two terminals (Supporting Information Figure S8). Based on this concept, we produced a measurement sample and deployed it to fulfill the cross section of a rectangular waveguide, as shown in Figure 2c (see the Experimental Section for more details of the measurement sample and method). Without any external bias, the structure exhibited minimal reflectance for both a short pulse and a CW at a frequency of 3.6 GHz with a 10 dBm input power level due to limited effective drain-source resistance (see $V_g$ = 0.0 V of Figure 2d and e). The current channel was then squeezed by decreasing the bias voltage supplied from a DC source so that, eventually, the difference in the reflectance reached more than 60 %, which corresponded to approximately 15 dB (see $V_g$ = -1.4 V of Figure 2d and e and Supporting Information Figure S12). More details, including the relationship between the effective drain-source resistance and bias voltage, are summarized in Supporting Information Figure S6 to Figure S11. Also, Supporting Information Figure S13 gives an additional example to tune transmittance instead of reflectance (or absorptance).

More complex tuning can be achieved by exploiting the electric potential difference varying across capacitors or inductors in the time domain. **Figure 3** illustrates the application of this concept to realize non-reciprocal scattering[14–17]. The periodic structure used in this study (see Figure 3a) resembles the waveform-selective absorbing metasurface introduced in Figure 2a, except that it has a perforation in the ground place and thus permits partial transmission. The amount of transmission can be controlled by the aperture size or alternatively by an additional capacitor if deployed across the aperture. Indeed, such a capacitor can be replaced with a pair of varactor diodes, whose parasitic capacitance varies in response to an applied voltage, as shown in Figure 3b. Under these circumstances, the transmittance peak can be shifted to a higher frequency by increasing the bias voltage, as plotted in the inset of Figure 3b. Note that this result is obtained by small signal simulations, wherein the schottky diodes connected to the top patches are not yet turned on. By increasing the input power level (to 25 dBm), however, the electric charges can be rectified and stored in the capacitor, which can be used to bias the pair of varactors through the use of an additional conducting via (see Figure 3a) and Supporting Information Figure S14). Such a waveform-selective metasurface was prepared as shown in Figure 3c and d and was tested inside a rectangular waveguide (see the Methods section for more details of the measurement sample and setup). With a 50-ns short pulse, which induces a limited amount of electric charges, the varactor diodes are only weakly biased, and the incident wave is not strongly transmitted at 2.7 GHz (Figure 3e). By contrast, much more power is transmitted using a CW at the same frequency, as the incident wave induces more electric charges, sufficiently biasing the pair of varactor diodes and shifting the transmission peak to a frequency close to 2.7 GHz. Interestingly, the difference between the short pulse and CW is reduced when the same waves are propagated upward through the unit cell (Figure 3f). In this case, presumably, the incoming waves induce fewer electric charges on the top patches as the standing wave distribution inside the substrate varies (Supporting Information Figure S16). However, this is not sufficient to achieve our non-reciprocal waveform-selective response, as the transmittance peak seen in the inset of Figure 3b remains at a low frequency. For this reason, the non-reciprocity and waveform selectivity disappear if the varactor diodes are replaced with fixed capacitors (3 pF) (Supporting Information Figure S17).

In addition to these fundamental scattering controls, waveform-selective metasurfaces potentially provide additional design flexibility in applied electromagnetic problems. **Figure 4**a and b describes a well-known resonant problem often seen in a conducting enclosure or cavity. Specifically, a strong external field can be excluded by entirely covering an object with conducting walls - for instance, to protect sensitive electronic devices. However, small openings or apertures often exist from the viewpoint of a realistic design (such as for ventilation). This leads to more strongly sensing external fields if the conductive enclosure resonates at specific frequencies $f_{m,n,p}$ [18]:

$$f_{m,n,p} = \frac{c_0}{2} \sqrt{\left(\frac{m}{a_x}\right)^2 + \left(\frac{n}{a_y}\right)^2 + \left(\frac{p}{a_z}\right)^2}, \quad (1)$$

where $c_0$ is the speed of light and $a_x$, $a_y$, and $a_z$ are the internal dimensions of the enclosure - set to 54, 54, and 36 mm here, respectively (Figure 4a). $m$, $n$, and $p$ are integers, no more than one of which can be zero. This equation clearly shows that resonance is correlated with the dimensions of the enclosure and is thus a physically inevitable issue.

One of the classical solutions to this issue is to design a part of the conducting walls with an absorbing material to dissipate the resonant energy[11,19], although this simultaneously prevents waves from transmitting through the apertures. As an alternative solution, this study replaces the front conducting wall (i.e., the one containing slits in Figure 4a) with a waveform-selective transmitting slit structure composed of inductor-based circuits (Figure 4c). Under these circumstances, a short pulse is not rectified by the diode bridges due to the electromotive force of the inductors. Hence, as usual, the slits resonate to permit the transmission of the incoming wave, resulting in poor shielding effectiveness close to or lower than 0 dB near 3.2 GHz (Figure 4d). In the same frequency region, however, CWs are largely shielded, as electric charges induced by this waveform can enter the diode bridges, which prevents the resonance of the slits. As a result of these features, this waveform-selective conducting enclosure enables control over the shielding effectiveness by varying the level of transmission through the apertures. Note that compared to the analytically derived resonant frequency of the enclosure (i.e., 3.9 GHz), the minimum shielding effectiveness for the pulse measurement appeared near 3.2 GHz in Figure 4d due to the presence of a monopole antenna that coupled to the resonance of the enclosure, although this antenna was still necessary for sensing the internal field.

The concept of waveform selectivities is not limited to the examples demonstrated above but could be extended to the design of other types of scattering parameters and even electromagnetic properties (e.g., permittivity and permeability) if the circuit components are properly deployed between conductor edges where a strong electric field is obtained, as seen in the insets of Figure 1e, h, and j. Consequently, the operation of the rectification process depends on the waveform of the incident wave. In this case, our idea is potentially applicable to recently studied fields, for instance, to perfect lenses[20,21], cloaking[22,23], and wavefront shaping[24,25]. However, this requires further improvements in waveform-selective performances with respect to the dynamic range and difference between pulse and CW responses. Furthermore, waveform-selective metasurfaces can be designed to respond not only to simple short and long pulses but also to intermediate pulses[10,26]. Note that on the one hand circuit values are involved with determining the time constant of waveform selectivity (i.e., transient response) but on the other hand these values need to be individually considered to achieve large difference between pulse and CW performances (see Supporting Information Figure S4). Importantly, our structures are planar periodic surfaces, similarly with FSSs[11], which are an old class of metamaterial[27,28] and metasurface[8]. However, compared to conventional FSSs, which only respond to waveforms of *different* frequencies, our structures additionally sense the difference in the waveform of an incoming wave even at the *same* frequency, which gives us a higher degree of control over transmission and reflection (compare Supporting Information Figure S5 to Figure 1e).

In conclusion, we have investigated the theoretical basis and demonstrated the experimental feasibility of circuit-based metasurfaces to control various types of fundamental electromagnetic scattering parameters in response to the waveform of the incoming wave. Waveform-selective metasurfaces were shown to be externally or internally tunable, which allowed the design of more complex scattering responses. In addition, the concept of waveform selectivity potentially provides additional design flexibility in applied electromagnetic issues ranging from traditional antenna design[29] to wireless communications[10] and even emerging areas such as perfect lenses[20,21], cloaking[22,23], and wavefront shaping[24,25].

**Experimental Section**

*Simulations*: Numerical simulations were performed using a co-simulation method that integrated an electromagnetic solver (Ansys HFSS) with a circuit simulator (Ansys Designer). In this method, the circuit components necessary for the design of waveform-selective metasurfaces were replaced with lumped ports in electromagnetic simulations. The actual components were then connected to the metasurfaces through the lumped ports in circuit simulations. This co-simulation method significantly reduced the time to sweep input power, frequency and other design parameters, reducing the total simulation time and facilitating optimization of the waveform-selective performance. Reflectance $R$, transmittance $T$, and absorptance $A$ for pulsed signals were calculated by *entirely* integrating reflected and transmitted waveforms in the time domain (specifically, from 0 to 100 ns for 50-ns pulses), which were then compared to the energy of the incident wave $E_i$, namely, $R = E_r/E_i$, $T = E_t/E_i$, and $A = 1 - R - T$, where $E_r$ and $E_t$, respectively, represent the energies of the reflected and transmitted waves. For CW simulations, incident waves continued until the time-domain response reached a steady state (mostly in 10 μs for this study). Subsequently, the incident, reflected, and transmitted energies were calculated by integrating their waveforms *only* for 10 ns, which was sufficiently long to ignore discretization errors in the time domain.

*Measurement Samples*: The measured waveform-selective metasurface samples used in Figure 2, Figure 3, and Figure 4 were prepared by using Rogers3003 as their substrates. They had periodic conducting patterns made of copper. Additionally, a thin coating layer was added to the surface of the conductors both to prevent oxidation and to facilitate soldering of the circuit components necessary for waveform selectivity. The dimensions of these samples are listed in Supporting Information Table S4, Table S6, and Table S8. In these samples, we used commercial schottky diodes, jFETs, and varactor diodes provided by Broadcom (HSMS-2860/2863/2864), Toshiba (2SK209 BL), and Skyworks (SMV2019-079LF), respectively. The conducting enclosure used in Figure 4 was assembled using the above-mentioned sample as well as copper plates coated for antioxidation.

*Measurement Methods*: For the measurements in Figure 2 and Figure 3, a signal generator (Anritsu MG3692C) was used as a signal source (see Figure 2c). This was connected to a standard rectangular waveguide (WR284) where a measurement sample was deployed. In the case of Figure 2, a circulator (Pasternack PE8401) was connected between the signal generator and waveguide so that the incident wave entered the waveguide, while the reflected wave was propagated to an oscilloscope (Keysight Technologies DSOX 6002A). This circulator was removed for the measurement shown in Figure 3, and the oscilloscope was connected to another side of the waveguide to measure the transmittance. A DC source (Keysight Technologies E3631A) was additionally introduced for the measurement shown in Figure 2 to bias jFETs via cables. In both measurements, the reflectance $R$, transmittance $T$, and absorptance $A$ were obtained similarly to simulations (see the Simulations subsection).

The measurement in Figure 4 used a vector network analyzer (VNA; Keysight Technologies N5249A) instead of the pair of the signal analyzer and oscilloscope (see Figure 4c). Note that pulse measurements were conducted using the pulse measurement mode of the VNA to measure performance for pulsed waves but not for CWs. The incident waves were radiated from a standard horn antenna (WR284) to a conducting enclosure (54 × 54 × 36 mm$^3$) including a waveform-selective metasurface. An 18-mm-tall monopole antenna was connected to the bottom of the enclosure to receive transmitted waves, which were eventually measured at the second port of the VNA via a coaxial cable. This transmittance $T_w$ was compared to that without the shielding walls (see Supporting Information Figure S19), $T_{wo}$, to obtain the shielding effectiveness *S.E.* as follows: *S.E.* = $10\log_{10}(T_{wo}/T_w)$ dB. In this measurement setup, we also deployed an amplifier, couplers, and attenuators to increase the input power level, maintain the calibration of the VNA and protect instruments from a large input power, respectively.

References

[1]   H. Chen, W. J. Padilla, J. M. O. Zide, A. C. Gossard, A. J. Taylor, R. D. Averitt, *Nature*. **2006**, *444*, 597.

[2]   A. Alu, G. D'Aguanno, N. Mattiucci, M. J. Bloemer, *Phys. Rev. Lett.* **2011**, *106*, 123902.

[3]   Y. Shen, D. Ye, I. Celanovic, S. G. Johnson, J. D. Joannopoulos, M. Soljačić, *Science*. **2014**, *343*, 1499.

[4]   Y. Shen, C. W. Hsu, Y. X. Yeng, J. D. Joannopoulos, M. Soljačić, *Appl. Phys. Rev.* **2016**, *3*, 11103.

[5]   H. Tao, A. C. Strikwerda, K. Fan, W. J. Padilla, X. Zhang, R. D. Averitt, *Phys. Rev. Lett.* **2009**, *103*, 147401.

[6]   T. Driscoll, H. T. Kim, B. G. Chae, B. J. Kim, Y. W. Lee, N. M. Jokerst, S. Palit, D. R. Smith, M. Di Ventra, D. N. Basov, *Science.* **2009**, *325*, 1518.

[7]   I. V Shadrivov, P. V Kapitanova, S. I. Maslovski, Y. S. Kivshar, *Phys. Rev. Lett.* **2012**, *109*, 83902.

[8]   D. Sievenpiper, L. Zhang, R. F. J. Broas, N. G. Alexópolous, E. Yablonovitch, *IEEE*

*Trans. Microw. Theory Tech.* **1999**, *47*, 2059.

[9] H. Wakatsuchi, S. Kim, J. J. Rushton, D. F. Sievenpiper, *Phys. Rev. Lett.* **2013**, *111*, 245501.

[10] H. Wakatsuchi, D. Anzai, J. J. Rushton, F. Gao, S. Kim, D. F. Sievenpiper, *Sci. Rep.* **2015**, *5*, 9639.

[11] B. A. Munk, *Frequency Selective Surfaces: Theory and Design*, A Wiley-Interscience Publication, New York, NY, **2000**.

[12] M. Born, E. Wolf, *Principles of Optics*, Cambridge University Press, Cambridge, U.K., **1999**.

[13] H. Wakatsuchi, C. Christopoulos, *Appl. Phys. Lett.* **2011**, *98*, 221105.

[14] Z. Wang, Z. Wang, J. Wang, B. Zhang, J. Huangfu, J. D. Joannopoulos, M. Soljačić, L. Ran, *Proc. Natl. Acad. Sci. USA*. **2012**, *109*, 13194.

[15] D. L. Sounas, C. Caloz, A. Alù, *Nat. Commun.* **2013**, *4*, 2407.

[16] Y. Shi, Z. Yu, S. Fan, *Nat. Photonics*. **2015**, *9*, 388.

[17] A. M. Mahmoud, A. R. Davoyan, N. Engheta, *Nat. Commun.* **2015**, *6*, 8359.

[18] C. Christopoulos, *Principles and Techniques of Electromagnetics Compatibility*, CRC Press Taylor & Francis Group, London, U.K., **2007**.

[19] H. Wakatsuchi, J. Paul, C. Christopoulos, *IEEE Trans. Antennas Propag.* **2012**, *60*, 5743.

[20] J. B. Pendry, *Phys. Rev. Lett.* **2000**, *85*, 3966.

[21] N. Fang, H. Lee, C. Sun, X. Zhang, *Science.* **2005**, *308*, 534.

[22] J. B. Pendry, D. Schurig, D. R. Smith, *Science.* **2006**, *312*, 1780.

[23] D. Schurig, J. J. Mock, B. J. Justice, S. A. Cummer, J. B. Pendry, A. F. Starr, D. R. Smith, *Science.* **2006**, *314*, 977.

[24] N. Yu, P. Genevet, M. A. Kats, F. Aieta, J. P. Tetienne, F. Capasso, Z. Gaburro, *Science.* **2011**, *334*, 333.

[25] N. Yu, F. Capasso, *Nat. Mater.* **2014**, *13*, 139.

[26] H. Wakatsuchi, *Sci. Rep.* **2015**, *5*, 16737.

[27] D. R. Smith, W. J. Padilla, D. C. Vier, S. C. Nemat-Nasser, S. Schultz, *Phys. Rev. Lett.* **2000**, *84*, 4184.

[28] R. A. Shelby, D. R. Smith, S. Schultz, *Science.* **2001**, *292*, 77.

[29] A. B. Constantine, *Antenna Theory: Analysis and Design*, John Wiley & Sons, **2005**.

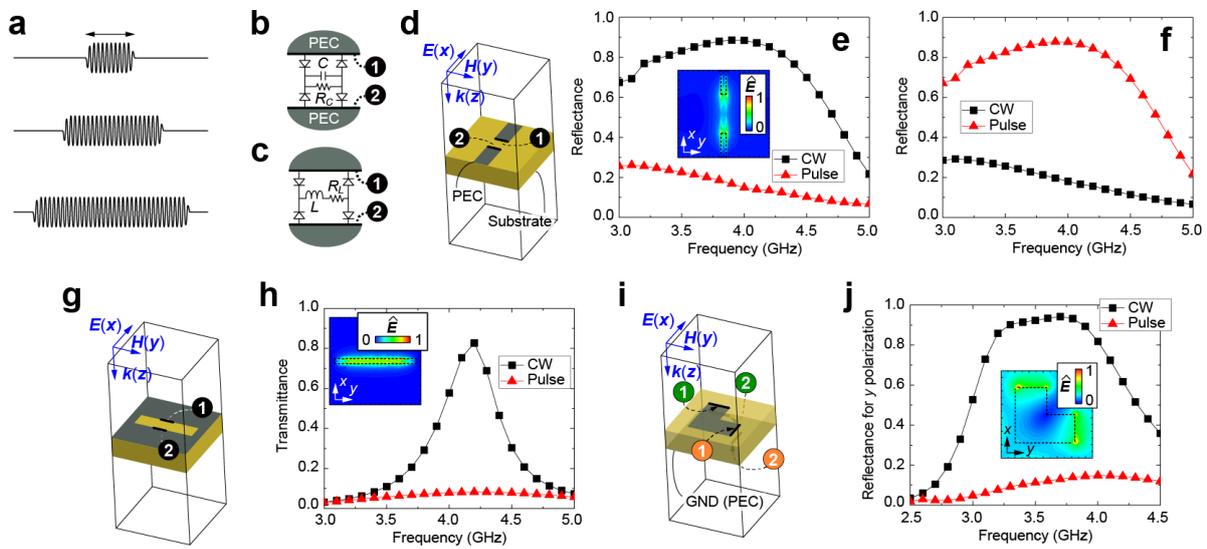

**Figure 1.** Fundamental scattering parameters controlled through waveform selectivity. a) Image of various pulse widths at the same frequency (see also Supporting Information Figure S1 for the relationship between the pulse width and spectrum). b,c) Capacitor- and inductor-based circuits deployed between conductors. The circuit values are given as $C$ = 1 nF, $L$ = 1 mH, $R_C$ = 10 kΩ, and $R_L$ = 31.2 Ω. The resonant frequencies of the capacitor and inductor are set to 300 MHz and 2.4 MHz, respectively. d) Periodic unit cell of the cut-wire structure to control reflection. e) With a capacitor-based circuit, the cut-wire structure shows a relatively strong level of reflection for a continuous wave (CW), while the reflecting performance is reduced for a 50-ns short pulse. f) With an inductor-based circuit, the structure enhances reflection for a short pulse, while a CW is weakly reflected. g) Periodic unit cell of the slit structure to control the transmission. h) With a capacitor-based circuit, the slit structure strongly transmits a CW. i,j) An L-shaped structure (containing capacitor-based circuits) reflects an incident wave with a polarization change depending on its waveform. Only the L-shaped structure contains two capacitor-based circuits per periodic unit cell. The insets of (e), (h), and (j) represent normalized electric field ($\hat{E}$) distributions calculated at the resonant frequencies of the small signal analysis (specifically, at 3.44 GHz, 4.07 GHz, and 3.15 GHz, respectively, as seen in Supporting Information Figure S3). The dashed lines draw each conductor shape, and these distributions were obtained either at the front surface of the structures (the cut-wire and slit structures) or at the middle of the substrate (the L-shaped structure).

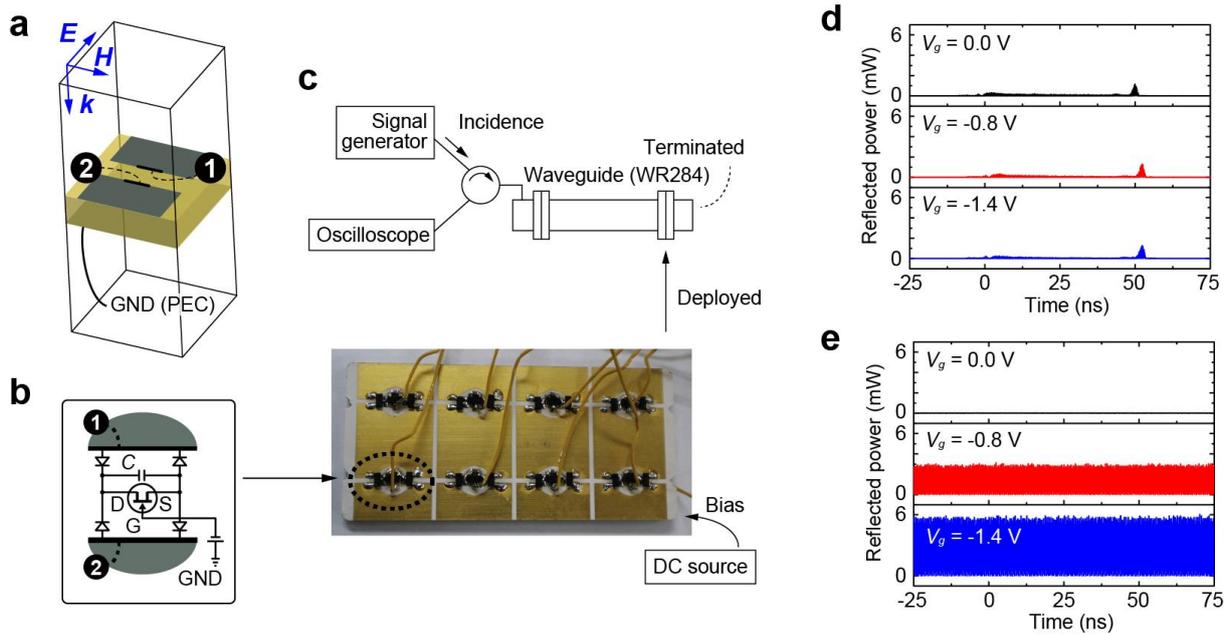

**Figure 2.** Waveform selectivity tuned by an external bias source. a) Periodic unit cell. b) The circuit configuration deployed between patches ($C$ = 1 nF). This resembles the capacitor-based circuit (i.e., Figure 1b), but the resistor is replaced with a jFET, which varies its effective resistance between the drain and source in relation to the gate voltage. c) Measurement setup. d,e) Experimentally measured reflected power for a short pulse and a CW signal with various gate voltages (see also Supporting Information Figure S11 for mode details). Without a gate bias, the difference between the pulse and CW cases is small; however, applying a gate voltage of -1.4 V leads to more than 60 % difference (or 15 dB on the decibel scale) in absorption (Supporting Information Figure S12). Consequently, waveform selectivity can be modulated between on and off depending on the electric stimulation.

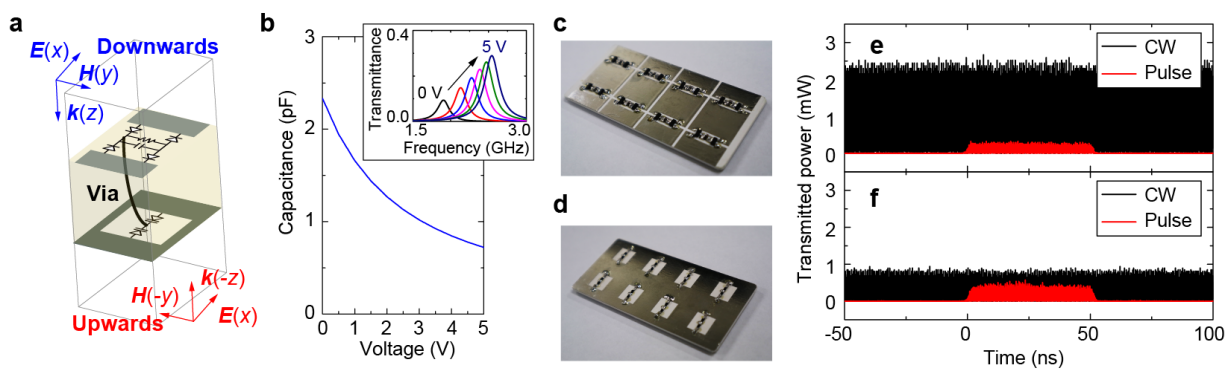

**Figure 3.** Waveform selectivity tuned internally to achieve non-reciprocal transmission. a) Periodic unit cell of a waveform-selective non-reciprocally transmitting metasurface. b) A varactor diode varies its parasitic capacitance in response to the bias voltage. The inset represents the transmittance of the metasurface in the small signal analysis. In this simulation, the conducting via is removed for the sake of simplicity, but the varactors are biased by a DC source, which shifts the peak of the maximum transmittance to a higher frequency. c,d) Front and back surfaces of the measurement sample. e,f) Experimental validation of waveform-

selective non-reciprocal transmission for downward and upward propagations, respectively (2.7 GHz, 25 dBm). The difference between the transmitted power levels of a short pulse and a CW is more clearly observed in the case of downward propagation.

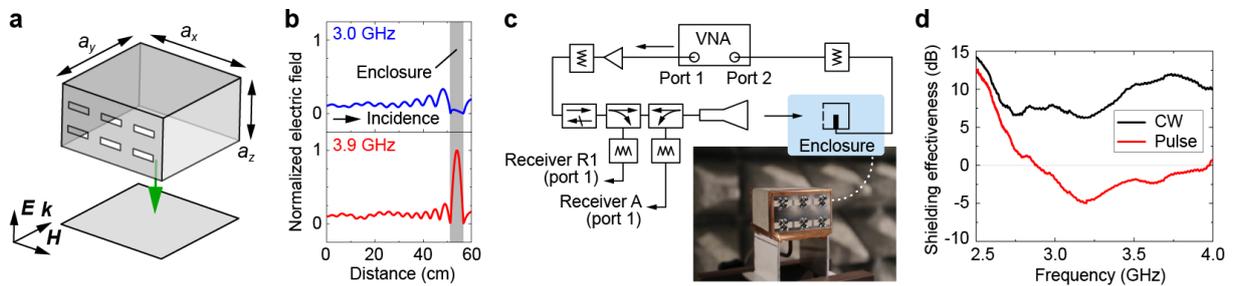

**Figure 4.** Waveform selectivity applied for electromagnetic issues to enhance device design flexibility. a) External fields can be separated with a conducting enclosure. b) The fields, however, appear more strongly at intrinsic resonant frequencies of the enclosure. c,d) Experimental setup and result using the conducting enclosure including a waveform-selective transmitting metasurface (see Supporting Information Figure S18 for the corresponding simulation). The use of waveform selectivity provides different levels of shielding effectiveness depending on the waveform so that, for instance, a short pulse can still be transmitted through the front apertures for wireless communications, while a CW at the same frequency would be blocked, and any sensitive electronics are therefore protected.

Hiroki Wakatsuchi,* Jiang Long, and Daniel F. Sievenpiper

**Title: Waveform Selective Surfaces**

Supporting Information

**Waveform Selective Surfaces**

*Hiroki Wakatsuchi,\* Jiang Long, and Daniel F. Sievenpiper*

- **Spectrum of various waveforms**
- **Cut-wire, slit, and L-shaped structures**
- **Waveform-selective metasurface using jFETs as the external tuning method**
- **Non-reciprocal waveform-selective metasurface**
- **Waveform-selective metasurface applied for the resonant conducting enclosure issue**

**Spectrum of various waveforms**

An incident wave oscillating at a single frequency may be represented by a cosine function $\cos(2\pi f t_0)$, where $f$ and $t_0$ denote frequency and time, respectively, and other variables (e.g., spatial coordinates, phase, and magnitude) are omitted for the sake of simplicity. If the wave experiences full wave rectification due to the presence of diode bridges, as in our circuit-based metasurfaces, the waveform is converted to $|\cos(2\pi f t_0)|$. When $x$ is set to these waveforms, the Fourier series expansions $g(x)$ are yielded as follows:

$$g(x) = \frac{1}{2}a_0 + \sum_{n=1}^{\infty} a_n \cos(nx) + \sum_{n=1}^{\infty} b_n \sin(nx), \quad (2)$$

where

$$a_0 = \frac{1}{\pi}\int_{-\pi}^{\pi} g(x)dx, \quad (3)$$

$$a_n = \frac{1}{\pi}\int_{-\pi}^{\pi} g(x)\cos(nx)dx, \quad (4)$$

$$b_n = \frac{1}{\pi}\int_{-\pi}^{\pi} g(x)\sin(nx)dx \quad (5)$$

and $n$ represents an integer. Using these equations, Table S1 shows the ratios of fields and energy for each frequency component. This table clearly indicates that after full wave rectification, most of the energy of the incoming wave is at zero frequency.

**Table S1.** Relative magnitudes of frequency components before and after full wave rectification

| Frequency | Before rectification | After rectification (in fields) | After rectification (in energy) |
|---|---|---|---|
| 0 | 0 | 2 | 4 |
| $f$ | 1 | 0 | 0 |
| $2f$ | 0 | 4/3 | 8/9 |
| $3f$ | 0 | 0 | 0 |
| $4f$ | 0 | 4/15 | 8/225 |
| $5f$ | 0 | 0 | 0 |
| ⋮ | ⋮ | ⋮ | ⋮ |

Next, we numerically show the spectrum of pulsed incident waves. A similar but analytical study is presented in our earlier work[26]. In this study, we use a circuit simulator (Ansys Designer R18.1) and simply oscillate a pulsed cosine wave with various widths at 3.9 GHz, which is then Fourier-transformed to obtain the spectra in Figure S1. This figure reveals that if the pulse is long enough (e.g., more than 20 ns), then the spectrum of the incident pulse appears around the oscillation frequency (3.9 GHz). However, reducing the pulse width to, for example, 1 ns leads to the generation of other non-negligible frequency components due to the discontinuity of the waveform at the beginning and end. If this extremely short pulse was used, our metasurfaces or even other ordinary metasurfaces would behave differently by sensing the difference in the frequency components. Importantly, however, our study used a pulse width of 50 ns, which ensures that our structures responded to the difference in the pulse width.

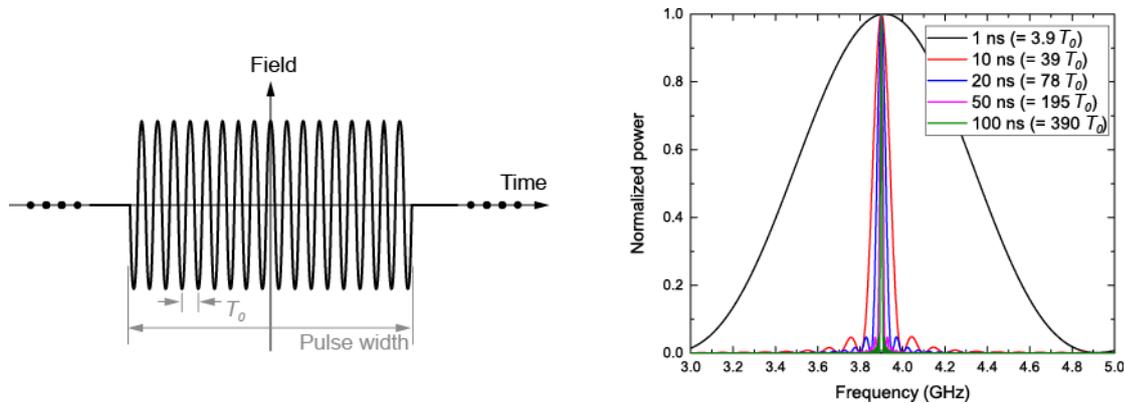

**Figure S1. Spectra of various pulse widths.** If the pulse is sufficiently long, then the spectrum appears around the oscillation frequency (3.9 GHz in this example).

**Cut-wire, slit, and L-shaped structures**
Figure 1 showed the waveform-selective scattering controls using cut-wire, slit, and L-shaped structures. The periodic units of these structures were designed as summarized in Figure S2 and Table S2. Periodic boundary conditions were applied to the directions of the incident $E$ and $H$ fields so that each model was effectively simulated as an infinite periodic array. The incident power was set to 10 dBm for the cut-wire and slit structures and 20 dBm for the L-shaped structure. These values were large enough to turn on the schottky diodes used (see their SPICE parameters in Table S3). As shown in Figure S3, which represents the results of a small signal analysis for these structures, the cut-wire and slit structures used in our study enhance the reflectance and transmittance, respectively, at designed frequencies as ordinary structures, while the L-shaped structure changes the polarization of the reflected wave. In addition, compared to Figure 1, the three structures exhibit slightly different results, as the diodes are turned on in Figure 1 (but not in Figure S3), which varies the frequency dependence of the scattering parameters due to the parasitic capacitance of the diodes.

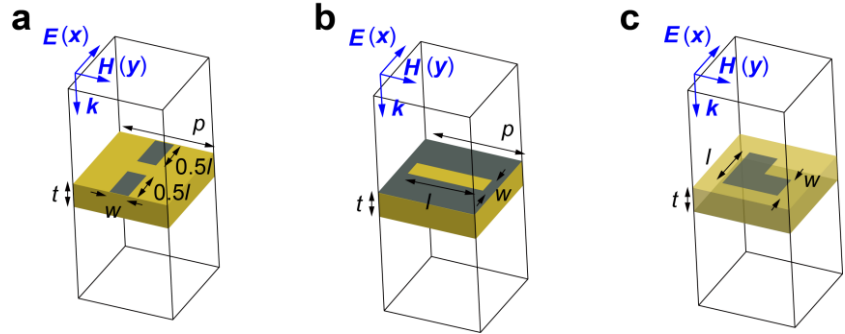

**Figure S2. Design of periodic units of cut-wire, slit, and L-shaped structures.** The design parameters are given in Table S2.

**Table S2.** Design parameters used for periodic unit cells of the waveform-selective metasurfaces in Figure S2. All of the numbers represent lengths in millimeters. All of the structures used Rogers3010 as their substrates.

| Design parameter | Cut-wire (Figure S2a) | Slit (Figure S2b) | L-shaped (Figure S2c) |
|---|---|---|---|
| $p$ | 18 | 18 | 13 |
| $l$ | 10 | 16 | 8 |
| $w$ | 1 | 1 | 4 |
| $t$ | 1.5 | 1.5 | 4.5 |

**Table S3.** SPICE parameters used for schottky diodes. These parameters were applied for all of the schottky diodes used in this study.

| SPICE parameter | Value |
|---|---|
| $B_V$ | 7.0 V |
| $C_{J0}$ | 0.18 pF |
| $E_G$ | 0.69 eV |
| $I_{BV}$ | $1 \times 10^{-5}$ A |
| $I_S$ | $5 \times 10^{-8}$ A |
| $N$ | 1.08 |
| $R_S$ | 6.0 Ω |
| $P_B$ (*VJ*) | 0.65 V |
| $P_T$ (*XTI*) | 2 |
| $M$ | 0.5 |

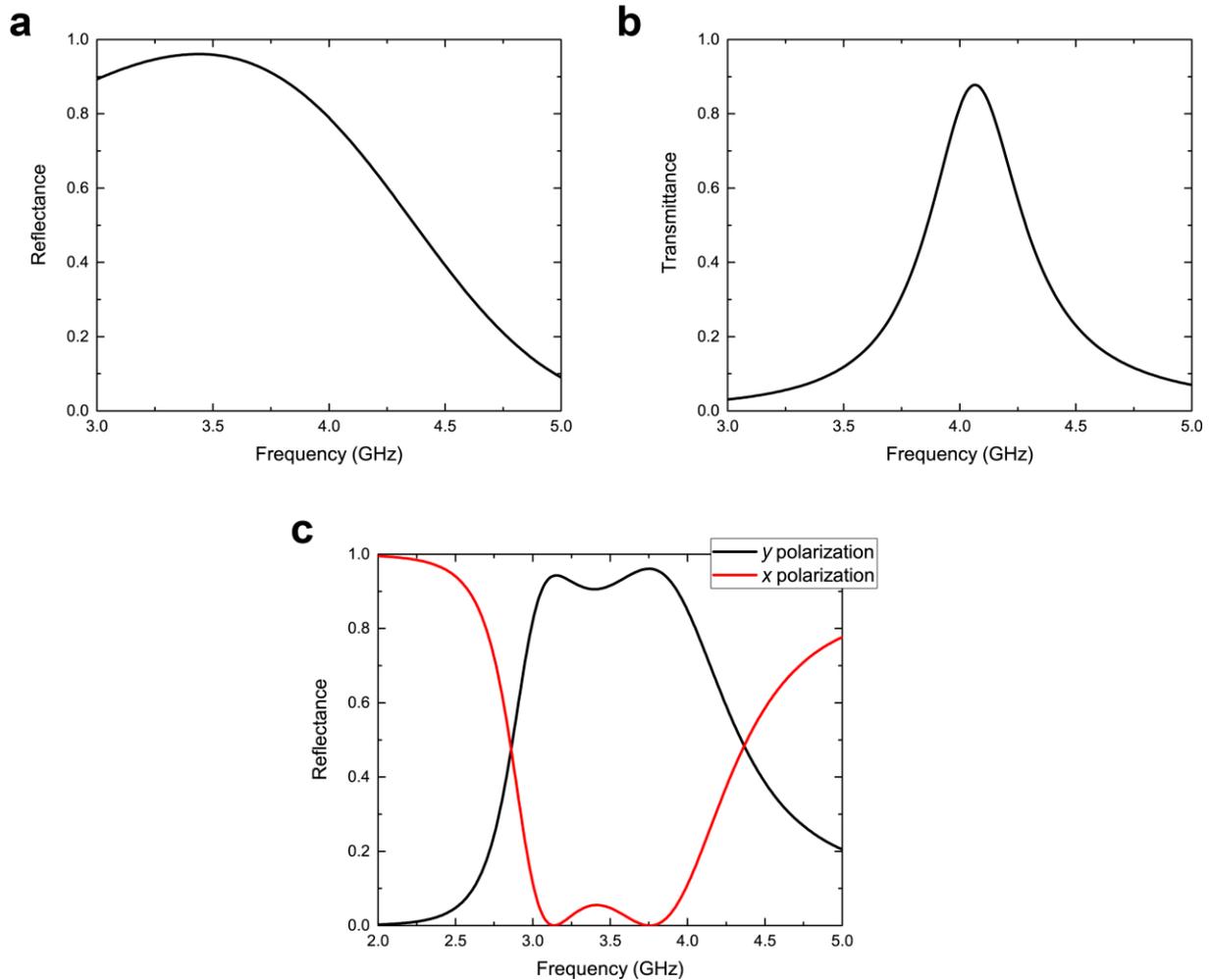

**Figure S3. Scattering parameters of cut-wire, slit, and L-shaped structures in small signal analysis. a,** Reflectance for the cut-wire structure. **b,** Transmittance for the slit structure. **c,** Reflectance for the L-shaped structure. These structures enhance each scattering parameter at the designed frequencies.

In addition, Figure S4 shows the reflectance of the cut-wire structure demonstrated in Figure 1e but with different values of $C$ and $R_C$. Note that by setting ($C$, $R_C$) to (1 nF, 10 kΩ) or (1 μF, 10 Ω), the product of $R_C$ and $C$ becomes the same value (i.e., $1.0×10^{-5}$ Ω·F), although their reflectances for a CW are different. This is because with a small resistance the rectified electric charges are allowed to come into the parallel resistor like the case of a pulse. Importantly, this indicates that the circuit values need to be individually considered to obtain waveform selectivity, although they also determine a time constant (i.e., transient response).

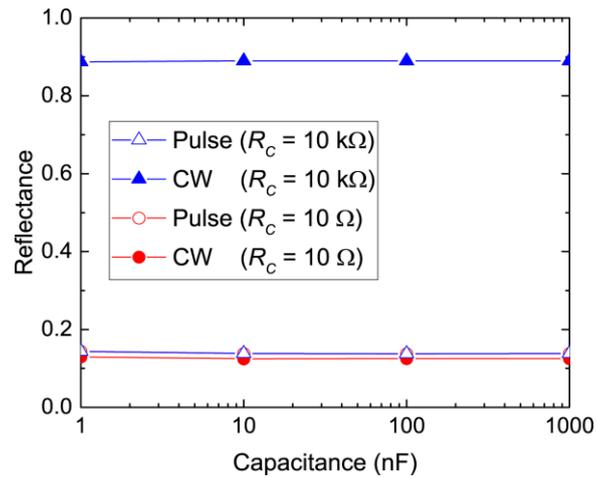

**Figure S4.** Reflectance of cut-wire structure (Figure 1e) with various values of $C$ and $R_C$.

Moreover, Figure S5 shows a simulation result of the cut-wire structure without all of the circuit components as an ordinary frequency selective surface (FSS). Under this circumstance, the cut-wire structure is unable to distinguish different waves at the same frequency. This numerical simulation clearly demonstrates the difference between conventional FSSs and our waveform-selective metasurfaces (cf. Figure 1e).

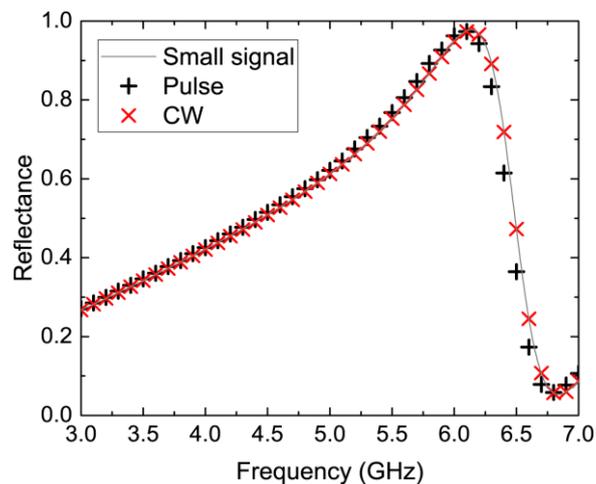

**Figure S5.** Reflectance of cut-wire structure without circuit components.

**Waveform-selective metasurface using jFETs as the external tuning method**

In Figure 2, we showed the effect of replacing the resistors of a metasurface with jFETs to realize a variable waveform-selective response. Without the jFETs (i.e., alternatively with 100 kΩ resistors), such a metasurface absorbs incoming waves as plotted in Figure S6a. In this simulation the input power level is set to 0 dBm. At 3.2 GHz, the transient reflected power varies as shown in Figure S6b, where the resistance value is swept from 10 Ω to 100 kΩ. Note that the transient reflected power significantly changes depending on the resistance, as it varies the amount of electric charges to be released for energy dissipation at the steady state.

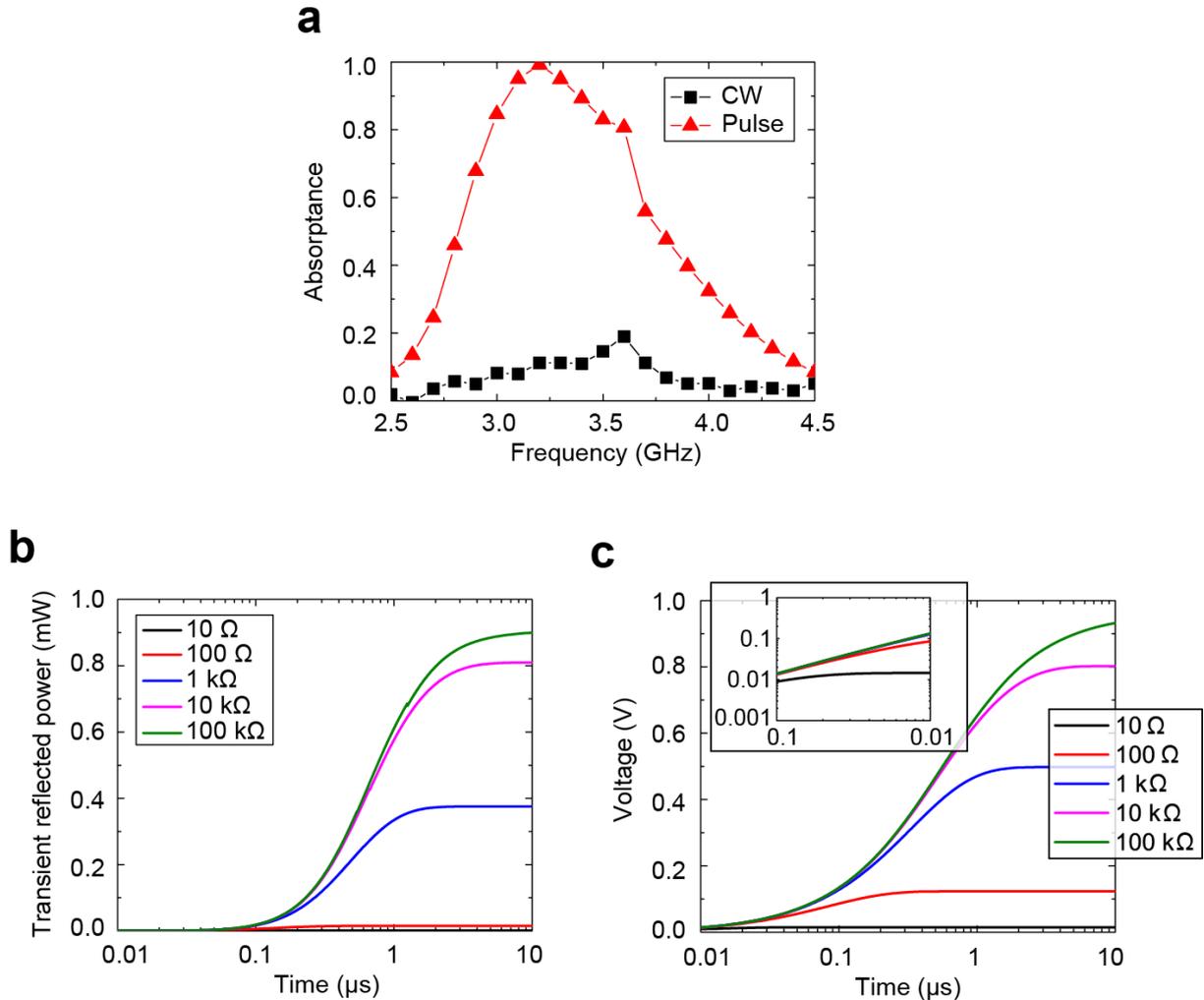

**Figure S6.** Response of the structure used in Figure 2a with the capacitor-based circuit (i.e., Figure 1b) instead of Figure 2b. **a,** Absorptance for short pulse and CW (0 dBm). $C$ and $R_C$ are set to 1 nF and 100 kΩ. **b,** Using various resistance values, the metasurface structure changes its transient average reflected power, namely, its absorption performance at 3.2 GHz. **c,** The corresponding capacitor voltage in each case.

In addition, this section shows the measurement results of the jFETs used in Figure 2 to clarify the relationship between the effective drain-source resistance and gate bias voltage. In this measurement, the drain-source current was measured by a source-measure unit (SMU; Keysight Technologies B2902A). By sweeping the gate bias voltage $V_g$ from 0.0 to -0.9 V, the drain-source current was found to decrease as the current channel was squeezed (Figure S7). Accordingly, the resultant effective drain-source resistance increased, as plotted in Figure S8. Importantly, the effective resistance depends not only on the gate voltage but also on the drain-source voltage, which, in our metasurface, was associated with the time-varying voltage built up across the capacitors. As a reference, we numerically estimated the voltage across the capacitor used in Figure S6b, as shown in Figure S6c. This revealed that the drain-source voltage varies within the range between 0.0 and 1.0 V, which indicates that the effective resistance changes in the time domain. Indeed, a small voltage appears even during the initial time period between 10 and 100 ns (Figure S6c inset). As shown in Figure S9, such a limited voltage or even a smaller value still allows the effective resistance to start at a large enough amount, for example, 1 kΩ if $V_g$ is set to -1.2 V. In other words, this effective resistance ensures that the response of the jFET-based metasurface does not remain unchanged like the curve of the 10 Ω data in Figure S6b. Notably, Figure S8 and Figure S9 are slightly different, which is simply due to the difference in the detectable current range of these jFET measurements. We also observed that these measurements used a DC source to bias the gate of the jFETs, while the metasurface demonstrated in Figure 2c to e used rectified electric charges as the bias source, which contained not only zero frequency but also other high frequency components.

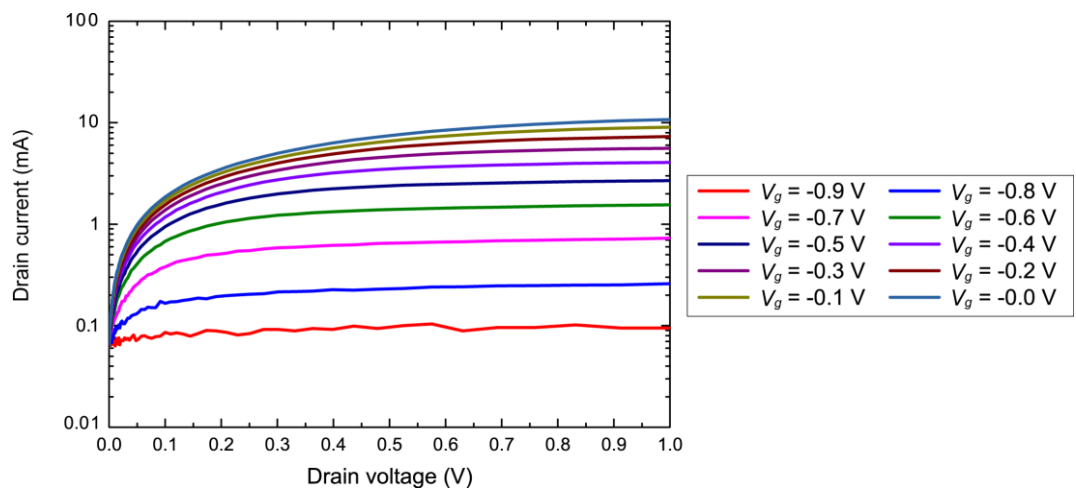

**Figure S7. Drain-source current with various gate voltages.** See also Figure S8 for the corresponding effective drain-source resistance.

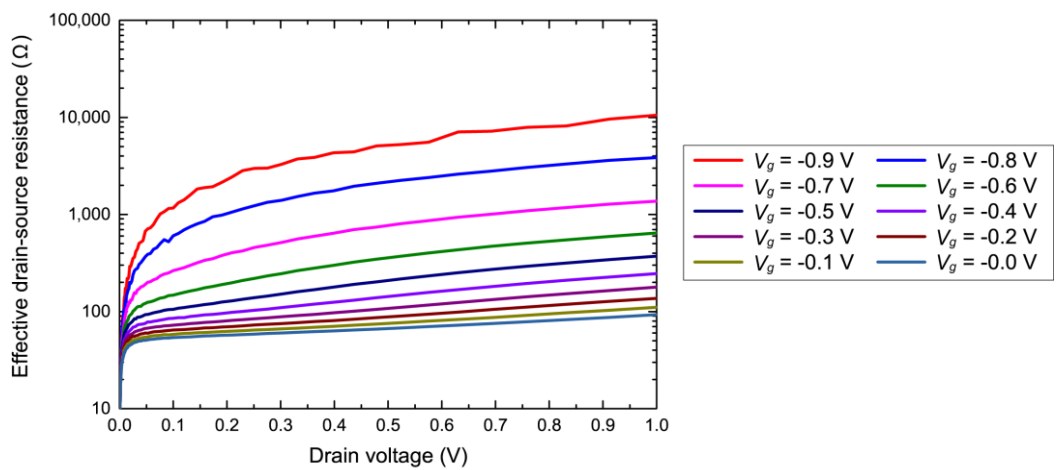

**Figure S8. Effective drain-source resistance with various gate voltages.** See also Figure S7 for the corresponding drain-source current and Figure S9 for the result on a smaller scale.

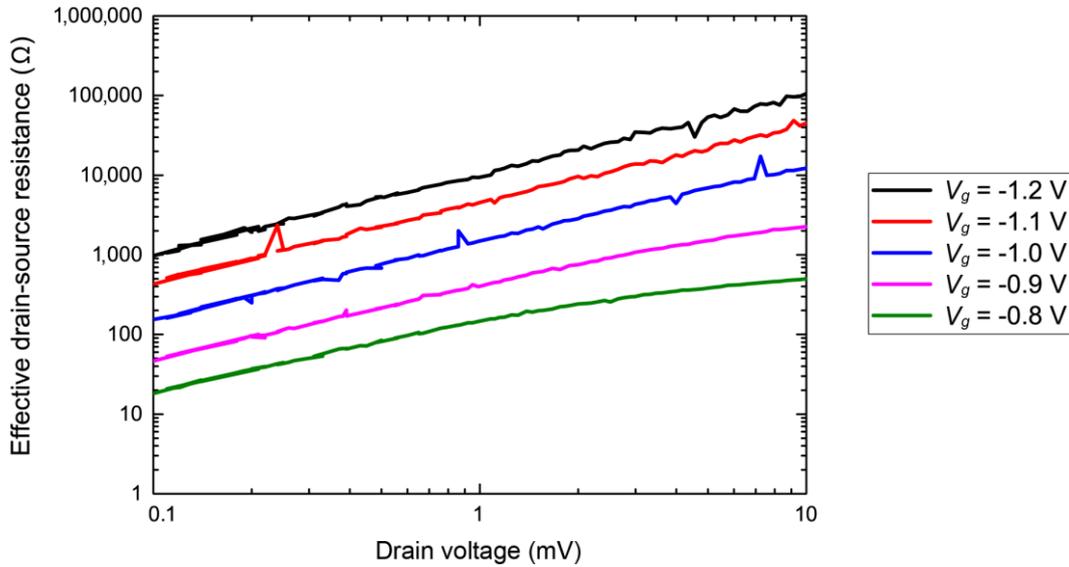

**Figure S9. Drain-source resistance with a small drain voltage.** See Figure S8 for the result on a larger scale.

The dimensions of the measurement sample used in Figure 2 are summarized in Figure S10 and Table S4. In this measurement, the input power level was set to 10 dBm, which was larger than the 0 dBm setting used in the simulations, which considers the energy density of the cross sectional area of the waveguide used (WR284). Figure S11 plots the reflected power shown in Figure 2d and e in finer steps, including the reference data using a copper plate instead. In this figure, the reflected power level of the copper plate was not exactly the same as the input power level (i.e., 10 dBm), as these waveforms were received by an oscilloscope after they experienced some loss factors including minor reflection and minor impedance mismatching at each component (see Figure 2c). However, by comparing these results to those of the metasurface for various pulse widths, the absorptance for the pulse was found to remain almost the same, while that for the CW varied, as plotted in Figure S12. Interestingly, this figure shows that the absorption performance for the CW becomes better than that for the pulse at $V_g = 0$ V. This is assumed to be due to the difference between their frequency spectra. As shown in Figure S1, the pulse width used in our study was set to be long enough so that the spectrum corresponded to the oscillation frequency. However, a minor part of the energy remained in other frequency components, which presumably appeared as the difference between the CW and pulse absorptances on the decibel scale in Figure S12.

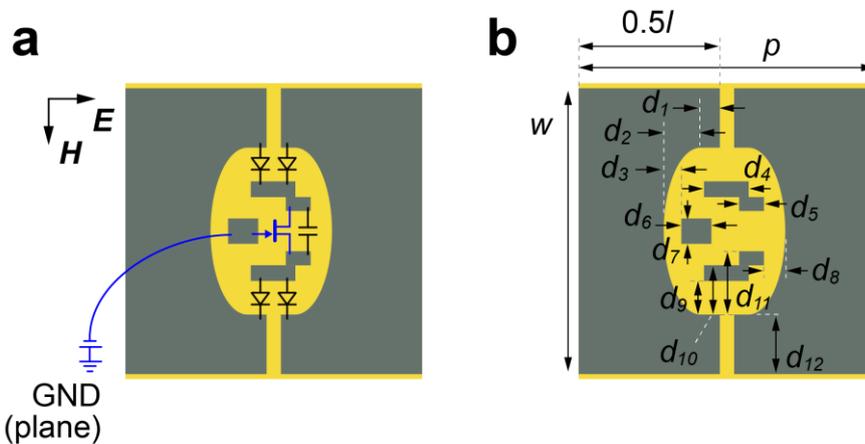

**Figure S10. Waveform-selective metasurface using jFETs. a,** Specific configuration of Figure 2a and b. **b,** Detailed dimensions without the circuit elements. Each design parameter is given in Table S4.

**Table S4.** Design parameters used for the periodic unit cell of the waveform-selective metasurface in Figure S10b. *t* represents the substrate thickness corresponding to Figure S2. The structure used Rogers3003 as its substrate.

| Design parameter | Value [mm] |
| --- | --- |
| $p$ | 17 |
| $l$ | 16 |
| $w$ | 16 |
| $t$ | 3 |
| $d_1$ | 0.7 |
| $d_2$ | 2.05 |
| $d_3$ | 1.35 |
| $d_4$ | 2.4 |
| $d_5$ | 1.2 |
| $d_6$ | 1.5 |
| $d_7$ | 1 |
| $d_8$ | 1.25 |
| $d_9$ | 1.3 |
| $d_{10}$ | 2.3 |
| $d_{11}$ | 3.3 |
| $d_{12}$ | 4.2 |

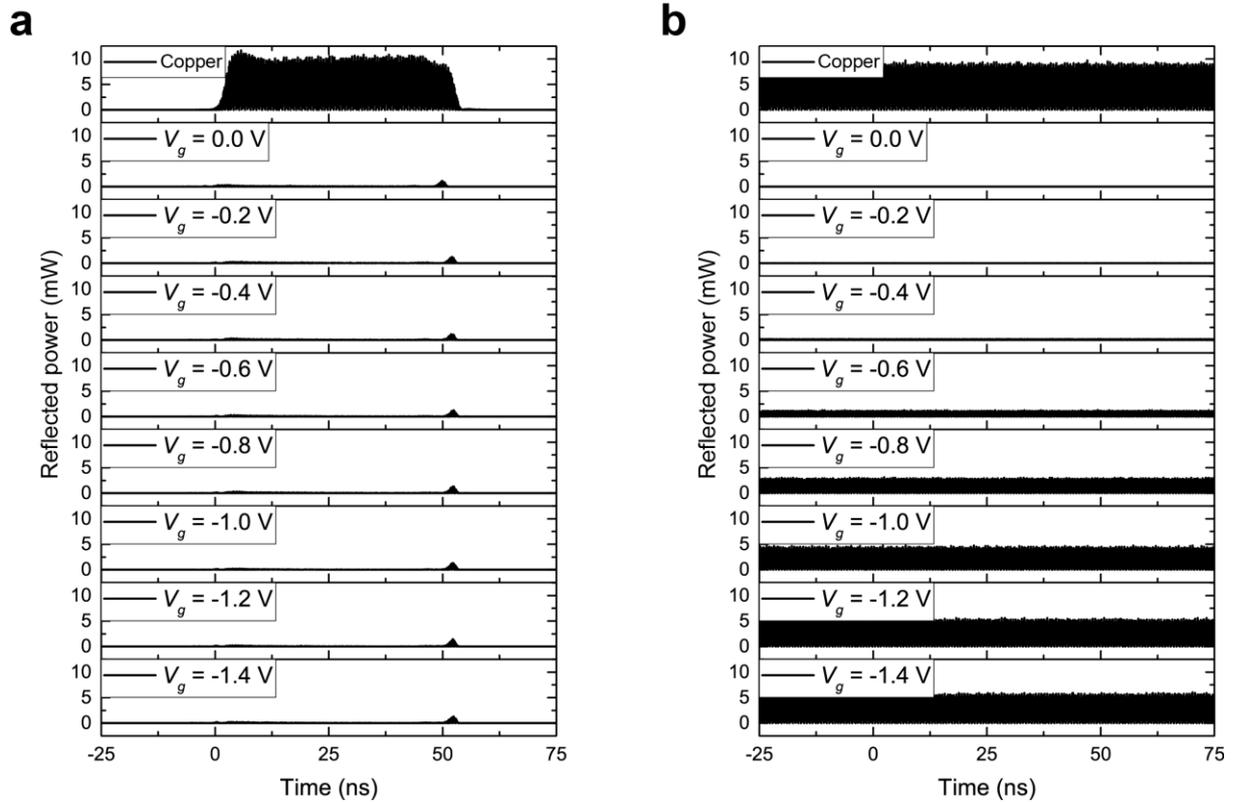

**Figure S11. Reflected power for various bias voltages in finer steps than those in Figure 2d and e. a,** Reflected power for a 50 ns short pulse. **b,** Reflected power for a CW. In the case of copper, the metasurface used in Figure 2c to e was replaced with a copper plate, which was assumed to have 100 % reflection in this study and was used to calculate the absorptances in Figure S12.

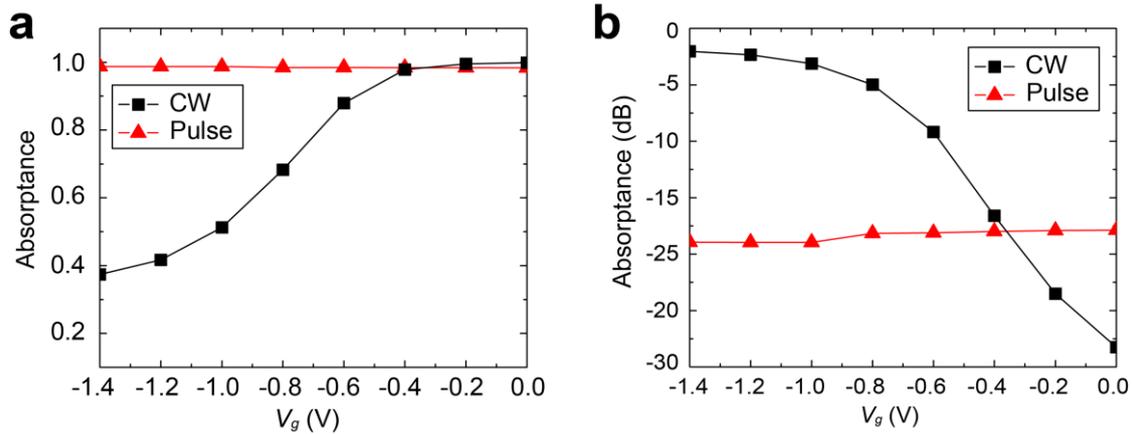

**Figure S12. Absorptance for various bias voltages. a,** Absorptance on the linear scale. **b,** Absorptance on the dB scale.

Additionally, Figure S13 demonstrates a tuning capability of a waveform-selective transmitting metasurface. Figure S13a to c shows details of the structure, which resembles the waveform-selective metasurface of Figure 2 and Figure S10. However, this waveform-selective transmitting metasurface was designed to fulfill the horizontal gaps on the front surface of the substrate (RO3010) and remove the ground plane from the back, which resulted in transmitting an incoming wave at a particular frequency. Besides, the direction of each diodes was changed as drawn in Figure S13a, because a strong electric field appears across the center of the vertical gap. Using the design parameters given in Figure S13b and Table S5, a measurement sample was fabricated to be 2 × 4 cells (Figure S13c) and tested inside a standard rectangular waveguide, similarly with the measurement in Figure 2. However, note that the oscilloscope seen in Figure 2 was connected to another side of the waveguide to observe a transmitted wave. In addition, the ground terminal of the DC source was electrically connected to a side wall of the waveguide, since theoretically an electric field becomes zero at this location. The frequency and power level of an incoming wave were set to 3.4 GHz and 15 dBm, respectively. Under these circumstances, the transmitted power was measured and normalized as plotted in Figure S13d and e. This figure clearly supports that the level of the transmitted power for a CW can be tuned by controlling the gate voltage $V_g$ of jFETs, as their effective resistance varies in response to $V_g$. On the other hand, the transmittance for a pulse remained the same. This tuning capability is similar with the one seen in Figure 2 but here controls transmittance instead of reflectance (or absorptance).

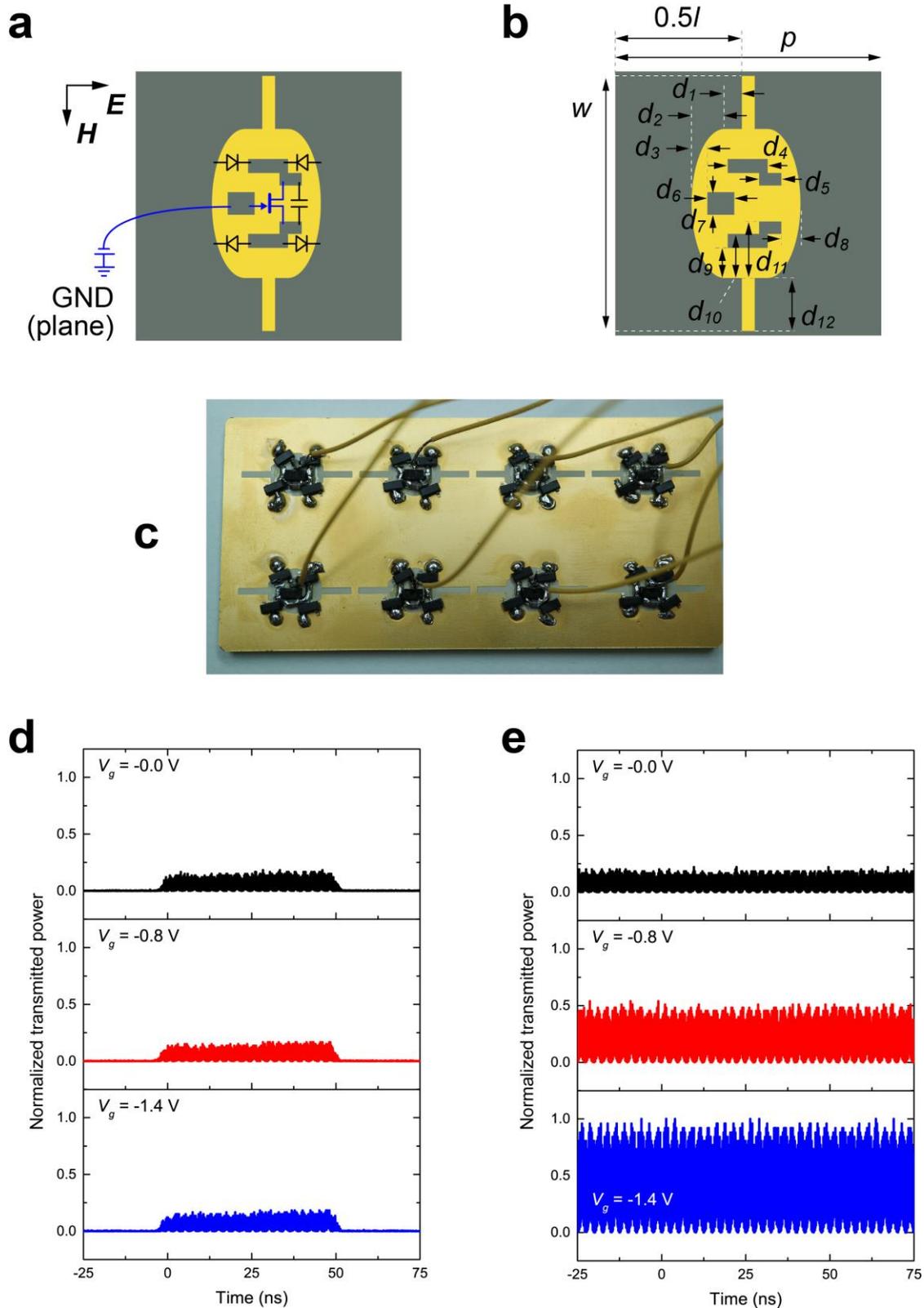

**Figure S13. Tuning capability of waveform-selective transmitting metasurface. a,** Specific configuration of the periodic unit cell used. The back side of this structure is not covered by a copper cladding to transmit an incoming wave. **b,** Detailed dimensions without the circuit elements. Each design parameter is given in Table S5. **c,** Measurement sample. **d,e,** Normalized transmitted power for a pulse and a CW.

**Table S5.** Design parameters used for the periodic unit cell of the waveform-selective transmitting metasurface in Figure S13b. *t* represents the substrate thickness corresponding to Figure S2. The structure used Rogers3010 as its substrate.

| Design parameter | Value [mm] |
|---|---|
| $p$ | 17 |
| $l$ | 16 |
| $w$ | 16 |
| $t$ | 2.5 |
| $d_1$ | 0.7 |
| $d_2$ | 2.05 |
| $d_3$ | 1.35 |
| $d_4$ | 2.4 |
| $d_5$ | 1.2 |
| $d_6$ | 1.5 |
| $d_7$ | 1 |
| $d_8$ | 1.25 |
| $d_9$ | 1.3 |
| $d_{10}$ | 2.3 |
| $d_{11}$ | 3.3 |
| $d_{12}$ | 4.2 |

**Non-reciprocal waveform-selective metasurface**

Figure 3 showed a non-reciprocal waveform-selective transmitting metasurface. A periodic unit cell of the measurement sample was designed, as shown in Figure S14 and Table S6. Figure S14 indicates that in the actual configuration, the metasurface contained additional capacitors connected to varactor diodes in series to adjust the amount of the total capacitance on the back side of the substrate (see $C_b$ in Figure S14b). The varactor diodes used to obtain the simulation results in Figure 3b were modeled using the SPICE model in Figure S15 and the design parameters in Table S7. Note that our structure varies the standing wave distribution inside the substrate, depending on the direction of the incoming wave, as reported by other group[17]. For instance, Figure S16 shows a simplified type of our structure and its corresponding standing wave distribution at 2.8 GHz. However, the difference in the standing wave distribution is not sufficient to obtain our non-reciprocal waveform-selective response, as the transmittance peak seen at a low frequency of the inset of Figure 3b is not shifted to the operating frequency. For this reason, both waveform-selective transmission and non-reciprocity disappeared if the varactor diodes were replaced with fixed capacitors (3 pF) (Figure S17). Note that this structure still contained all the other circuit components including schottky diodes.

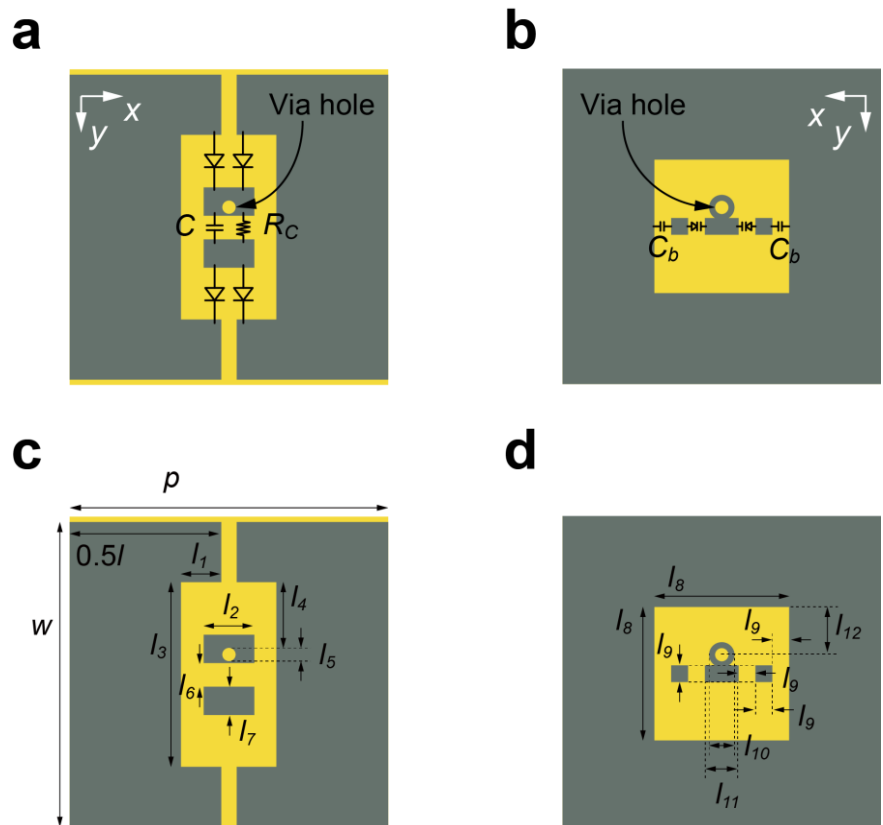

**Figure S14. Non-reciprocal waveform-selective metasurface. a,b,** Specific configurations on the top and bottom surfaces of its periodic unit cell. $C$, $C_b$, and $R_C$ were set to 1 nF, 3 pF, and 10 kΩ, respectively. **c,d,** Detailed geometries of the conducting areas. These design parameters are seen in Table S6.

**Table S6.** Design parameters used for the periodic unit cell of the non-reciprocal waveform-selective metasurface in Figure S14. $t$ represents the substrate thickness corresponding to Figure S2. The structure used Rogers3003 as its substrate.

| Design parameter | Value [mm] |
|---|---|
| $p$ | 17 |
| $l$ | 16 |
| $w$ | 16 |
| $t$ | 1.5 |
| $l_1$ | 1.7 |
| $l_2$ | 2.4 |
| $l_3$ | 7.6 |
| $l_4$ | 2.7 |
| $l_5$ | 0.3 |
| $l_6$ | 1 |
| $l_7$ | 2 |
| $l_8$ | 8 |
| $l_9$ | 1 |
| $l_{10}$ | 1.1 |
| $l_{11}$ | 2 |

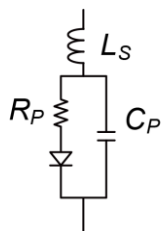

**Figure S15. SPICE model for varactor diodes.** This diode model was simulated using the parameters shown in Table S7.

**Table S7.** SPICE parameters used for varactor diodes. See also Figure S15.

| SPICE parameter | Value |
| --- | --- |
| $I_S$ | $1\times10^{-14}$ |
| $R_S$ | 0 |
| $N$ | 1 |
| $T_T$ | 0 |
| $C_{J0}$ | 2.25 pF |
| $M$ | 1.4 |
| $V_J$ | 3.5 V |
| $E_G$ | 1.11 |
| $X_{TI}$ | 3 |
| $K_F$ | 0 |
| $A_F$ | 1 |
| $F_C$ | 0.5 |
| $B_V$ | 0 |
| $I_{BV}$ | $1\times10^{-3}$ |
| $I_{SR}$ | 0 |
| $I_{KF}$ | 0 |
| $T_{NOM}$ | 27 |
| $L_S$ | 0.7 nH |
| $R_P$ | 4.8 Ω |
| $C_P$ | 0.07 pF |

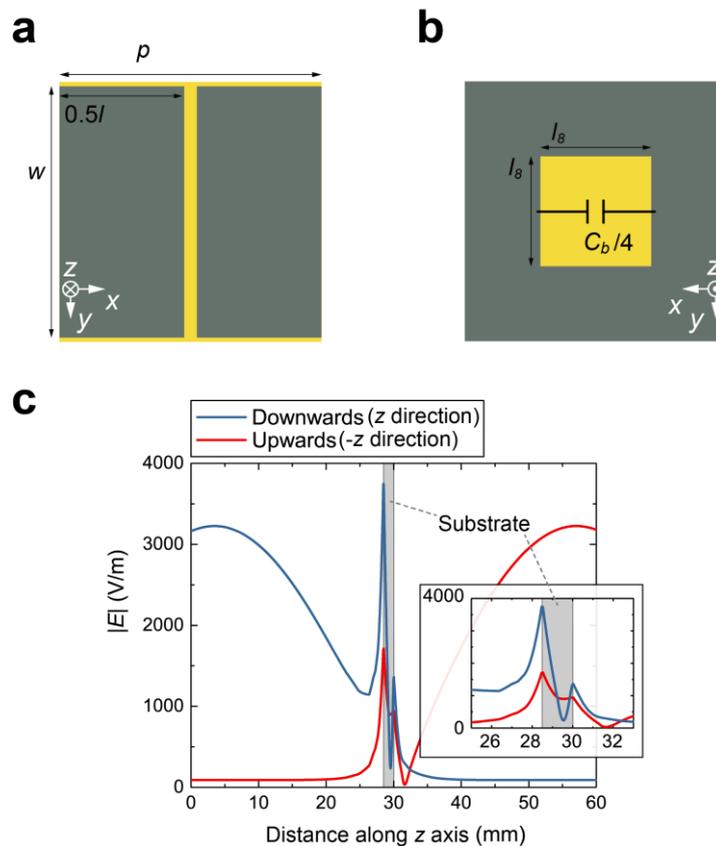

**Figure S16. Standing wave distribution of a simplified structure. a,b,** The simulation model (cf. Figure S14a,b). The design parameters were the same as the ones given in Table S6. $C_b/4$ was set to 0.75 pF to almost match the entire capacitance of the structure of Figure S14 in the small signal analysis (see also Figure 3b). **c,** Simulation result.

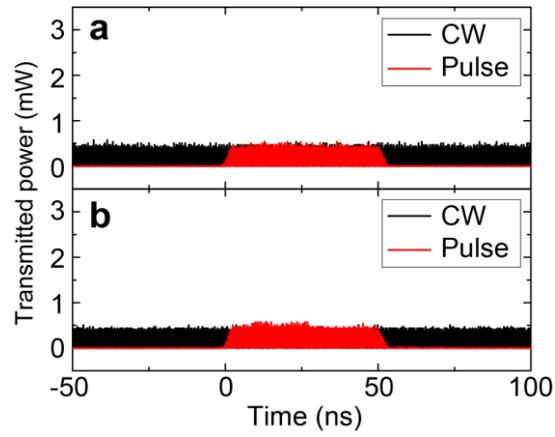

**Figure S17. Transmitted power of the structure used in Figure 3 without varactor diodes (cf. Figure 3e,f).**
**a,** Transmitted power for downward propagation. **b,** Transmitted power for upward propagation. In this structure, varactor diodes were replaced with fixed capacitors (3 pF).

**Waveform-selective metasurface applied for the resonant conducting enclosure issue**

In Figure 4, we experimentally demonstrated a waveform-selective metasurface with a conducting enclosure to control its shielding effectiveness depending on the incoming waveforms. In Figure S18, we performed an effectively similar simulation where an incident wave was launched from a horn antenna to a conducting enclosure including a waveform-selective metasurface. Table S8 shows the dimensions used in the simulation. Although the simulation result (i.e., Figure S18e) slightly differed from the measurement result (Figure 4d) due to minor differences between the simulation and measurement, including imperfection in the preparation of the measurement sample, the shielding effectiveness still varied in response to the type of the incoming waveform. As a reference, Figure S19 shows the measured enclosure without shielding walls, namely, a bare monopole antenna.

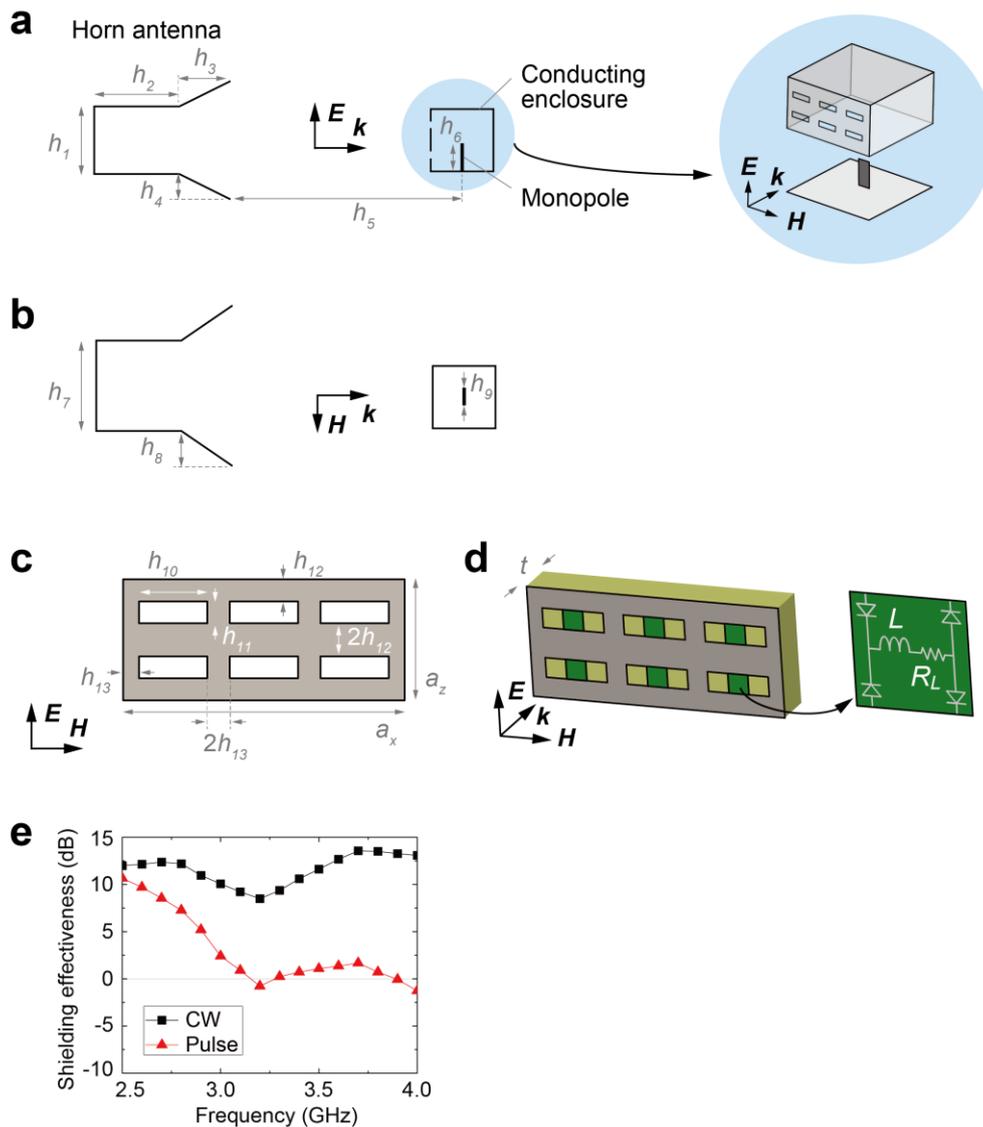

**Figure S18. Simulation of a resonant conducting enclosure using waveform-selective metasurface. a, b,** Entire configuration. An incident wave was set to propagate from a horn antenna to a monopole antenna connected to a conducting ground plane. The input power level was 35 dBm at the horn antenna, and each parameter value used is shown in Table S8. **c, d,** Front panel of the enclosure (i.e., waveform-selective transmitting metasurface). $L$ and $R_L$ were, respectively, set to 100 μH and 5.5 Ω. **e,** Simulation result of shielding effectiveness (cf. Figure 4d).

**Table S8.** Design parameters used for Figure S18. See Figure 4 for $a_y$. These dimensions were based on those used for the measurement (Figure 4) except for the monopole antenna deployed inside the conducting enclosure.

| Design parameter | Value [mm] |
|:---:|:---:|
| $h_1$ | 34 |
| $h_2$ | 141 |
| $h_3$ | 168 |
| $h_4$ | 34 |
| $h_5$ | 372 |
| $h_6$ | 18 |
| $h_7$ | 72 |
| $h_8$ | 35 |
| $h_9$ | 2 |
| $h_{10}$ | 16 |
| $h_{11}$ | 5 |
| $h_{12}$ | 6.5 |
| $h_{13}$ | 1 |
| $a_x$ | 54 |
| $a_y$ | 54 |
| $a_z$ | 36 |
| $t$ | 1.5 |

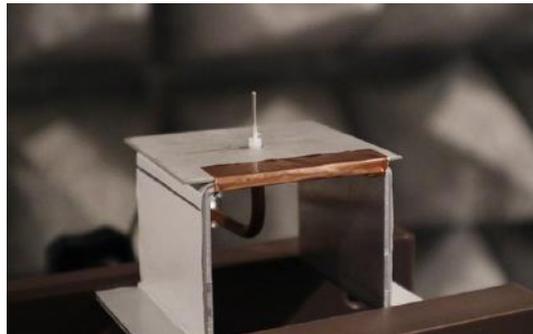

**Figure S19.** Bare monopole antenna (i.e., conducting enclosure without shielding walls and a waveform-selective metasurface).